\documentclass[aps,prl,twocolumn,amsmath,amssymb,superscriptaddress]{revtex4}
\usepackage{graphicx}
\usepackage{amsmath}

\begin{document}

\title{Interaction Induced Hall Response in a Spin-Orbit Coupled Bose-Einstein Condensate}
\author{Zhu Chen}
\affiliation{Institute for Advanced Study, Tsinghua University, Beijing, 100084, China}
\author{Hui Zhai}
\affiliation{Institute for Advanced Study, Tsinghua University, Beijing, 100084, China}
\date{\today}
\begin{abstract}
In this letter we consider the dynamic behaviors of spin-orbit coupled Bose condensates realized in recent experiments. We show that there exists an interaction induced ac Hall response which is absent in a non-interacting system. This condensate has two distinct equilibrium phases known as the plane wave phase and the stripe phase. In the plane wave phase, we show that an ac longitudinal current will induce an ac radial current in the transverse direction, and vice versa, as a cooperation effect of spin-velocity locking and spin-dependent interaction. In the stripe phase, we show that the dominant longitudinal response to a transverse radial current is sliding of the density stripe, because it is the low-lying excitation mode originated from spontaneous spatial translational symmetry breaking in this phase.
\end{abstract}
\maketitle

The Hall effect is known as the production of current transverse to the voltage difference, or vice versa. Conventionally, it is caused by the Lorenz force for particles in a magnetic field or the Coriolis force in a rotating frame, which couples the particle's motion in one direction to its motion in the transverse direction. In systems of neutral atomic gas with synthetic magnetic field \cite{magnetic,Bloch}, a superfluid Hall effect can manifest itself in the collective oscillations and has also been observed in a recent experiment \cite{Hall}.

With similar Raman coupling technique, an artificial spin-orbit (SO) coupling can also been generated in ultracold bosons \cite{NIST, shanxi, Shuai} and fermions \cite{Jing}.  As an analogy to condensed matter systems, conventional Hall and spin Hall effects have also been predicated for some SO coupled cold atom systems \cite{Zhu,Duine,Hu_Hall, Zhang}. However, since the configuration of SO coupling realized in current experiments is a special one as an equal weight combination of Rashba-type and Dresselhaus-type. Only in one spatial direction, where two Raman beams are counter-propagating, the motions of atoms are coupled to their spins. Thus, its single particle Hamiltonian along three different directions are separable and commute with each other. That means the motions along different directions are completely uncorrelated if no interactions. Therefore the conventional Hall and spin Hall responses are absent here.

Nevertheless, realizing SO coupling in cold atom systems brings several new ingredients that are absent in condensed matter systems. First, the system can be bosonic; secondly, the interactions between atoms are spin-dependent; and thirdly, there exists a harmonic confinement potential. These ingredients do give rise to intriguing physics for equilibrium physics, such as
stripe superfluid phase \cite{Zhai,SObecHo} and half vortex phase \cite{Wu,Victor,Hu,Santos}. In this letter we focus on whether they will also give rise to non-trival dynamics which has not been discovered in condensed matter systems before, and indeed we find that there exists an alternative and more striking Hall response arisen from spin-dependent interactions.

\begin{figure}[tbp]
\includegraphics[height=1.9in, width=3.4 in]
{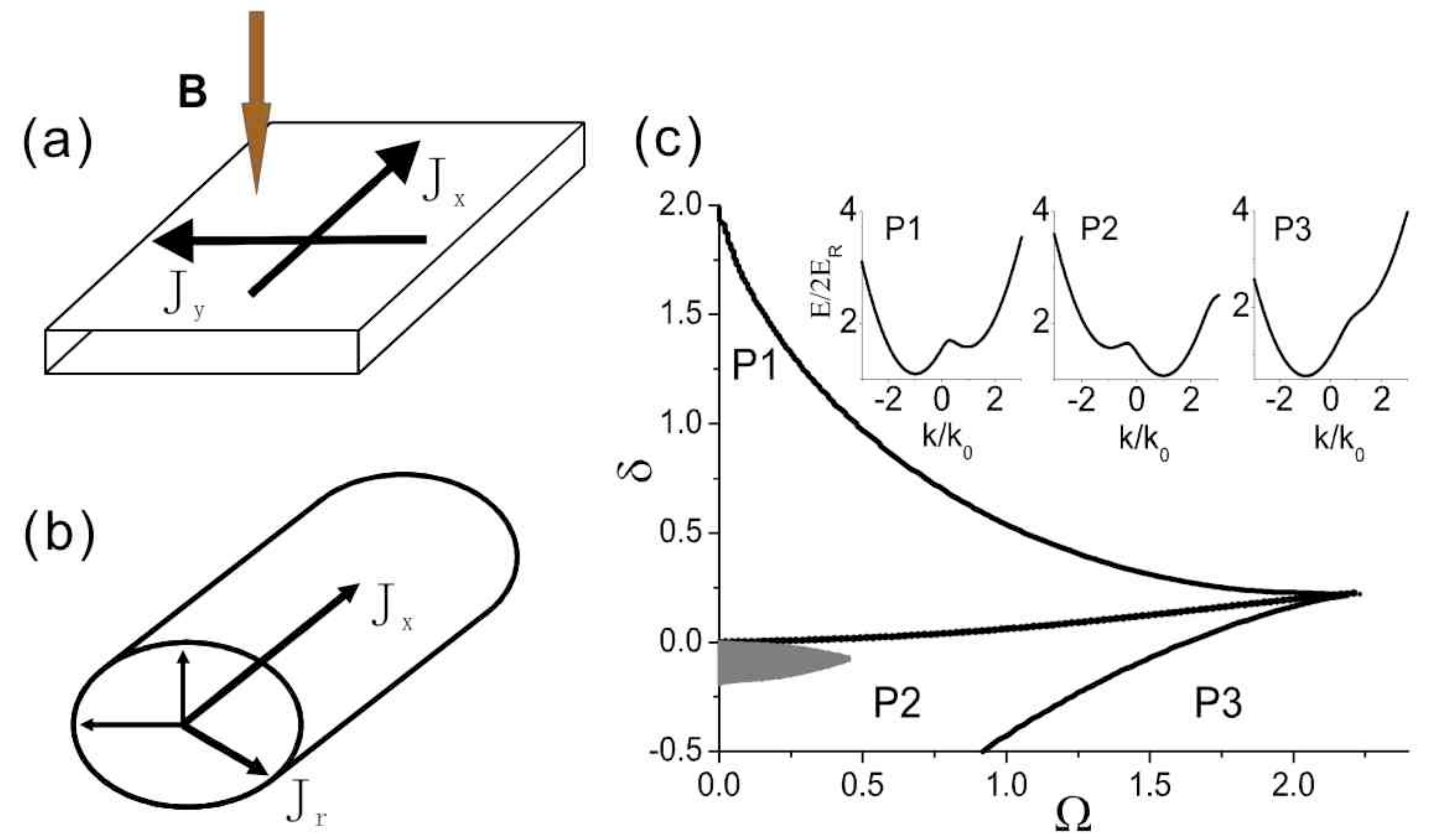}
\caption{(a) Configuration of conventional Hall effect; (b) Configuration of  Hall response proposed in this work; (c) Phase diagram for the system considered in this work. In P1 and P2 regimes the single particle dispersion has two local minima, while in P3 regime there is only one minimum. The grey regime is the stripe phase while its outside regime is the plane wave phase. $\Omega$ and $\delta$ are in unit of 2$E_{\text{R}}=\hbar^2 k^2_0/m$. In all plots of this work, we take $g_2=-g_0/10$. \label{Fig1}}
\end{figure}

The geometry under consideration is also slightly different from conventional ones. Conventional geometry of Hall response is shown in Fig. \ref{Fig1}(a), while in this system the Hall response is that an ac longitudinal current will induce an ac radial current in the transverse plane, as shown in Fig. \ref{Fig1}(b). For a trapped cloud, this is manifested as the coupling between the radial breathing mode and the longitudinal dynamics.

This dynamic behavior also strongly depends on the underlying equilibrium quantum phases and reveals their fundamental properties. It is known that there are two distinct superfluid phases in this system depending on external parameters, which are known the plane wave phase and the stripe phase \cite{NIST, SObecHo, LiYun}, as shown in Fig. \ref{Fig1}(c). In the plane wave phase, we show that an ac longitudinal current couples to transverse motion. This comes from that the SO coupling locks spin to the velocity. While in the stripe phase, the dominant response is the coupling between the transverse current and the longitudinal sliding mode of density stripe. This reveals the spontaneous spatial translational symmetry breaking in the stripe phase and the sliding mode is a low-lying Goldstone mode in this phase.

{\it Model.} We illustrate the physics using $F=1$ condensate as an example. Three spin states of $F=1$ are coupled by two counter-propagating laser beams along $\hat{x}$ direction \cite{NIST,shanxi,Shuai}, and the Hamiltonian of this system is
\begin{equation}
\mathcal{\hat{H}}=\int \left[\psi^\dag({\bf r})\mathcal{H}_0\psi({\bf r})+g_0\hat{n}^2({\bf r})+g_2\hat{S}^2({\bf r})\right]d^3{\bf r}
\end{equation}
where $\psi^\dag({\bf r})=(\psi^\dag_{-1}({\bf r}),\psi^\dag_0({\bf r}),\psi^\dag_{1}({\bf r}))$, $\mathcal{H}_0=\mathcal{H}_{0,x}+\mathcal{H}_{0,\perp}+V_{\text{trap}}$, and $\mathcal{H}_{0,x}$ is given by
\begin{align}
\begin{pmatrix} \frac{\hbar^2(k_x+k_0)^2}{2m}-\frac{\delta}{2} & \frac{\Omega}{2}  & 0 \\ \frac{\Omega}{2} & \frac{\hbar^2(k_x-k_0)^2}{2m}+\frac{\delta}{2} &  \frac{\Omega}{2} \\ 0 & \frac{\Omega}{2} & \frac{\hbar^2(k_x-3k_0)^2}{2m}+\frac{3\delta}{2}+\varepsilon \end{pmatrix}, \label{H0x}
\end{align}
$\mathcal{H}_{0,\perp}=\hbar^2k^2_{\perp}/(2m){\bf I}$, where $\perp$ denotes $(y,z)$ plane and ${\bf I}$ is $3\times 3$ identity matrix.
Here $2\hbar k_0$ is the momentum transfer during the Raman process, $\delta$ is two-photon Raman detuning, $\varepsilon$ is the quadratic
Zeeman energy.
$\hat{n}({\bf r})=\psi^\dag({\bf r})\psi({\bf r})$ and $\hat{{\bf S}}({\bf r})=\psi^\dag({\bf r}){\bf S}\psi({\bf r})$, where ${\bf S}$ is $3\times 3$ Pauli matrix vector. The energy dispersion for the lowest eigen-state of $\hat{H}_{0,x}$ is denoted by $\mathcal{\epsilon}(k_x)$ and is shown in the inset of Fig. \ref{Fig1}(c). $\mathcal{\epsilon}(k_x)$ has two local minima for small $\delta$ and small $\Omega$, and displays only one minimum for either large $\delta$ or large $\Omega$.
We consider a harmonic trap of cylindrical symmetry with $\omega_y=\omega_z=\omega_{\perp}$ and $V_{\text{trap}}=(m\omega_x x^2+m\omega_{\perp} r^2_{\perp})/2$. Thus, the single particle Hamiltonian is separable in three different directions and they commute with each other.

{\it Variational Wave Function Approach.} A variational wave function approach has been successfully used for studying dynamics of a BEC in absence of SO coupling \cite{Zoller}. Here we generalize it to the case with SO coupling. The variational wave function is taken as a product of an overall Gaussian profile and the spin wave function $\phi$.
\begin{equation}
\Phi=\prod\limits_{\eta=x,y,x}\left(\frac{2}{\pi R^2_\eta} \right )^{\frac{1}{4}}e^{-(\frac{1}{R^2_\eta}-\frac{i\delta_\eta}{2})(r_\eta-A_\eta)^2}\phi
\end{equation}
where ${\bf A}=(A_x,A_y,A_z)$ is the center-of-mass position, $R_{\eta}$ ($\eta=x,y,z$) denote the width of condensate, and $\delta_{\eta}$ is introduced as conjugate variable to $R_{\eta}$ \cite{Zoller}.
In the regime where $\mathcal{\epsilon}(k_x)$ has two minima at $k_{\pm}$, $\phi$ is assumed to be
\begin{equation}
\phi=\cos\theta e^{i\frac{\varphi}{2}}e^{i{\bf p}_+({\bf r-A})}\xi(p_{+,x})+\sin\theta e^{-i\frac{\varphi}{2}}e^{i{\bf p}_-({\bf r-A})}\xi(p_{-,x})\label{stripe}
\end{equation}
where ${\bf p}_{\pm}$ are momenta around energy minima $(k_{\pm},0,0)$, respectively. While in the regime where $\mathcal{\epsilon}(k_x)$ has a single minimum, saying at $k_+$, $\phi$ is assumed to be \begin{equation}
\phi=e^{i{\bf p}_+({\bf r-A})}\xi(p_{+,x})\label{PW}
\end{equation}
$\xi(p_{\pm,x})$ is a three-component normalized spin wave function taken as the lowest eigenstate of Hamiltonian Eq. (\ref{H0x}). This is the main assumption we make here, that is, the spin wave function is always locked to the center-of-mass velocity during the whole dynamic process due to SO coupling. This assumption is supported by recent experimental observations \cite{Shuai}.

To compute the energy with wave function Eq. (\ref{stripe}), we note typically $(p_{+,x}-p_{-,x})R_x\gg 1$, thus the terms that contain $e^{-(p_{+,x}-p_{-,x})^2R^2_x}$ will be ignored. This means that to a very good approximation the energy functional maintains spatial translational invariance, because the potential gradient due to the harmonic trap is much weaker compared with SO coupling. Under this approximation, the condensate energy $\mathcal{E}$ is given by
\begin{widetext}
\begin{equation}
\mathcal{E}=\sum\limits_{\eta=x,y,z}\left[\frac{1}{2R^2_\eta}+\frac{\delta^2_\eta R^2_\eta}{8}+\frac{\omega^2_\eta}{2}\left(A^2_\eta+\frac{R^2_\eta}{4}\right)\right] +\cos^2\theta\left(\epsilon(p_{+,x})+\frac{\hbar^2 p^2_{+,\perp}}{2m}\right) +\sin^2\theta\left(\epsilon(p_{-,x})+\frac{\hbar^2 p^2_{-,\perp}}{2m}\right)+\frac{f(\theta,p_{+,x},p_{-,x})}{\sqrt{\pi}^3R_xR_yR_z} \label{energy}
\end{equation}
\end{widetext}
where $f(\theta,p_{+,x},p_{-,x})$ denotes the interaction energy density of a uniform system
\begin{equation}
f(\theta,p_{+,x},p_{-,x})=\int d^3{\bf r}\left(g_0|\phi|^4+g_2|\phi^*{\bf S}\phi|^2\right)
\end{equation}
For the equilibrium state, obviously one has $\delta_\eta=0$, $A_{\eta}=0$ for $\eta=x,y,z$ and $p_{\pm,\perp}=0$. $p_{\pm,x}$, $\theta$ and $R_{\eta}$ should be determined by coupled equations $\partial \mathcal{E}/\partial p_{\pm}=0$, $\partial \mathcal{E}/\partial R_\eta=0$ and $\partial \mathcal{E}/\partial \theta=0$. If the solution gives $0<\theta<\pi/2$, the condensate wave function is a superposition which gives rise to the stripe order, and this phase is named as ``stripe superfluid" phase \cite{Zhai,SObecHo}, as indicated by the grey area of Fig. \ref{Fig1}(c). Note that $f$ is independent of $\varphi$ due to the translational symmetry, which means that $\varphi$ of an equilibrium state can take any value determined by spontaneous symmetry breaking. If the solution gives $\theta=0$ or $\theta=\pi/2$, or if $\mathcal{\epsilon}(k_x)$ has a single minimum as wave function Eq. (\ref{PW}), the system will be in a ``plane wave superfluid" phase. The main part of this paper below is to use this variational approach to study collective modes in these two phases, respectively.

{\it Plane Wave Phase:}  For the plane wave phase, without loss of generality, we can take $\theta=0$ in Eq. (\ref{stripe}) in the two minima regime or equivalently take Eq. (\ref{PW}) in the single minimum regime, and the energy $\mathcal{E}$ is given by Eq. (\ref{energy}) with $\theta=0$, and automatically $f$ depends on $p_{+,x}$ only. Considering the Lagrangian $\mathcal{L}=\int d^3{\bf r}\Phi^*(i\partial/\partial t-\hat{H})\Phi$, we arrive at
\begin{align}
\mathcal{L}&=\sum\limits_{\eta}\left(p_{+,\eta}\dot{A}_{\eta}-\frac{R^2_\eta\dot{\delta}_\eta}{8}\right)-\mathcal{E},
\end{align}
The equation-of-motion for dipole motion along $\hat{x}$ direction is given by
\begin{align}
\dot{A}_x=\frac{\partial \epsilon(p_{+,x})}{\partial p_{+,x}}+\frac{\partial f/\partial p_{+,x}}{\sqrt{\pi^3}R_xR_yR_z}; \    \  \dot{p}_{+,x}=-\omega^2_x A_x. \label{dipole1}
\end{align}
And the radial dynamics of breathing mode is given by
\begin{align}
&\dot{R}_{\eta}=R_{\eta}\delta_{\eta}\nonumber\\
&\dot{\delta}_{\eta}=\frac{4}{R^4_\eta}-\omega^2_{\eta}-\delta^2_\eta+\frac{4f}{\sqrt{\pi^3}R^2_{\eta}R_xR_yR_z} \label{breathing}
\end{align}

\begin{figure}[tbp]
\includegraphics[height=3.4in, width=3.3 in]
{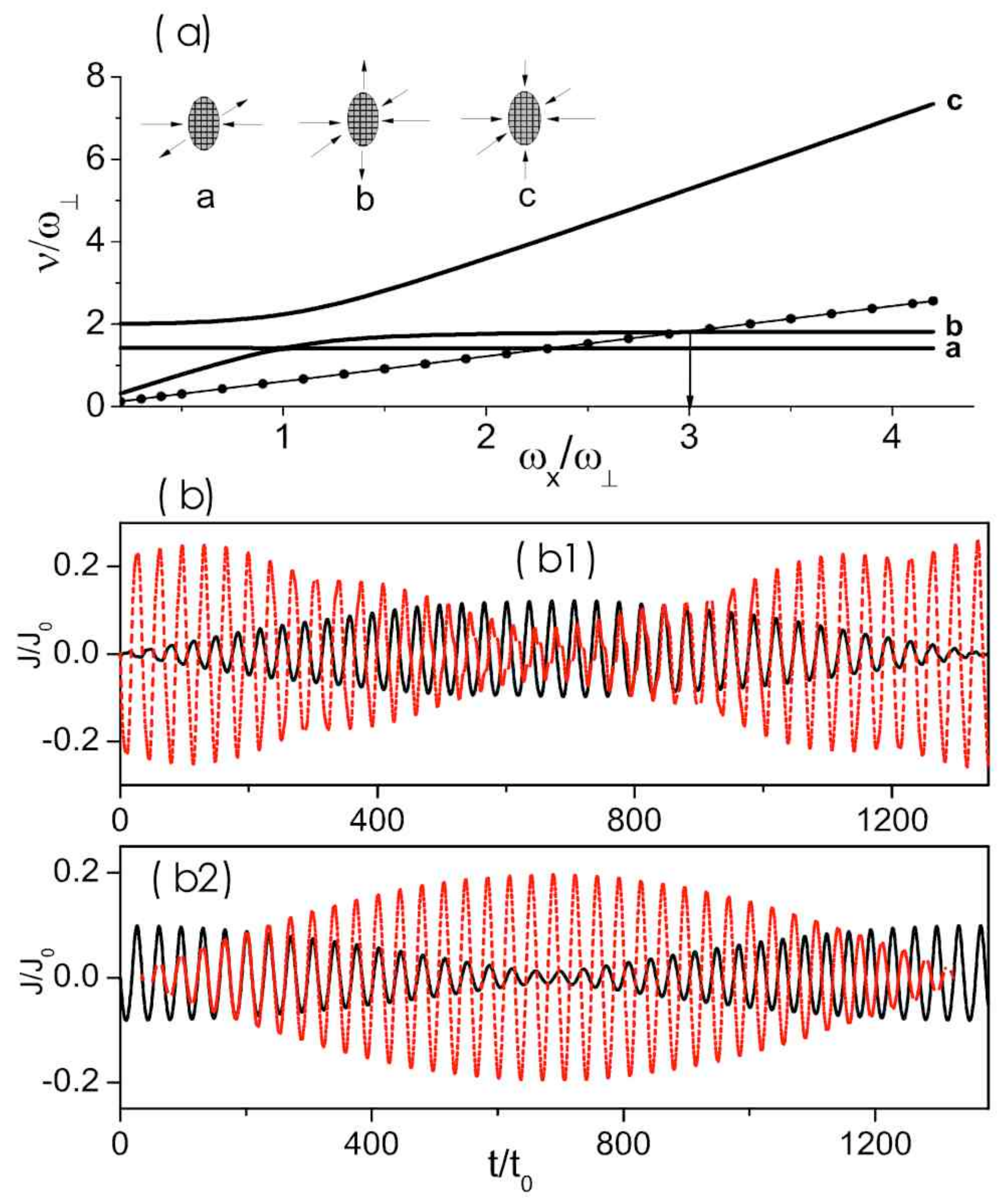}
\caption{Hall response in the plane wave phase. (a) Frequency of three breathing modes (solid lines labelled by (a), (b) and (c)) and the frequency of dipole mode (dots) as a function of $\omega_x/\omega_\perp$ in absence of spin-dependent interaction $g_2$. The arrow indicates the position of mode resonance. (b): Hall effect nearby resonance point. (b1): A longitudinal current (black solid lines) induced by excitation of transverse radial current (red dashed line); (b2): A transverse radial current (red dashed line) induced by excitation of longitudinal current (black solid line). $J$ is in unit of $J_0=\hbar k_0/m$. $t$ is in unit of $t_0=\hbar/2E_{\text{R}}$, and $E_{\text{R}}=\hbar^2 k^2_0/(2m)$. For this plot $\hbar\omega_x= 0.64 E_{\text{R}}$ and $\hbar\omega= 0.2 E_{\text{R}}$. For instance, in recent $^{87}$Rb experiment, $E_{\text{R}}=2\times 10^3$Hz and $t_0$ is $0.04$ms.\label{Hall}}
\end{figure}

The $f$ function contains both $g_0$ term and $g_2$ term. In the plane wave phase, the density is a constant, and thus the $g_0$ term is also a constant. The momentum dependence of $f$ comes from the $g_2$ term, because the spin configuration changes with momentum due to SO coupling and thus the spin-dependent interaction energy also changes. If $g_2$ is much smaller than $g_0$ and one can ignore $g_2$ term, then $\partial f/\partial p_{+,x}$ is zero, the dipole mode Eq. (\ref{dipole1}) and the breathing mode Eq. (\ref{breathing}) are completely decoupled. In this case, the breathing mode is the same as that in a scalar condensation derived in Ref. \cite{Zoller}, and the frequencies of three different breathing modes are shown in Fig. \ref{Hall}(a) using solid lines. While in contrast to a conventional condensate, the Kohn theorem will be violated due to SO coupling and the dipole mode frequency will be different from $\omega_x$ as shown in Fig. \ref{Hall}(a) using solid lines with dots \cite{Shuai}. In this case, the coupled equation Eq. (\ref{dipole1}) with $\partial f/\partial p_{+,x}=0$ predicts how the dipole mode frequency depends on $\delta$ and $\Omega$, which is very consistent with recent experiment on $^{87}$Rb \cite{Shuai}, where $g_2$ is only $0.46\%$ of $g_0$.

The coupling between dipole and breathing modes will become significant with larger $g_2$, for instance, if similar experiment is done with magnetic atoms like Cr \cite{Yi}, Dy \cite{Ben} and Er \cite{Ferlaino}. Nevertheless, even for smaller $g_2$, since there exists a window of aspect ratio $\omega_x/\omega_{\perp}$ in which the frequency of bare dipole mode is at resonance with one of the breathing modes, as indicated by arrow in Fig. \ref{Hall}(a), significant coupling between these two modes can be obtained near the resonance regime. In this case, the full response behavior requires numerically solving the coupled differential equations Eq. (\ref{dipole1}) and Eq. (\ref{breathing}). The results are illustrated in Fig. \ref{Hall}(b).

Fig. \ref{Hall}(b) illustrates the coupling between $J_{x}=\dot{A}_x$ and $J_{\perp}=\frac{-i\hbar}{2m}\int\left( \Phi^*\frac{\partial}{\partial r_{\perp}}\Phi-\Phi \frac{\partial}{\partial r_{\perp}}\Phi^* \right )$. In Fig. b1 we show when we excite a radial ac current $J_{\perp}$ by suddenly compressing the condensate in the transverse direction, $J_{\perp}$ will induce a longitudinal current $J_{x}$. Vice versa, in Fig. b2, we show when we excite a longitudinal ac current $J_x$ by suddenly displacing the condensate away from the center of harmonic trap, $J_x$ will induce a radial current $J_{\perp}$. This is precisely the Hall response induced by interactions. We also notice that there are two different frequencies in the dynamics. This is because in resonant regime, two nearly degenerate modes are split by spin-dependent interaction $g_2$ term. Thus the dynamics exhibits similar behavior of quantum beats. The larger frequency (smaller period) is nearly bare frequency, while the smaller frequency (larger period) is mostly determined by coupling provided by $g_2$.

{\it Stripe Phase:} Now we turn to the dynamics in the stripe phase. Here the dynamic equations involve both $A_{\pm}$ and $p_{\pm}$. The Lagrangian is given by
\begin{align}
\mathcal{L}&=\sum\limits_{\eta}\left(p_{c,\eta}\dot{A}_{\eta}-\frac{R^2_\eta\dot{\delta}_\eta}{8}\right)-\cos2\theta\frac{\dot{\varphi}}{2}-\mathcal{E},
\end{align}
where $p_{c,\eta}=\cos^2\theta p_{+,\eta}+\sin^2\theta p_{-,\eta}$. Thus, the equation for dipole mode is modified as
\begin{align}
&\dot{A}_x=\cos^2\theta\frac{\partial \epsilon(p_{+,x})}{\partial p_{c,x}}+\sin^2\theta\frac{\partial \epsilon(p_{-,x})}{\partial p_{c,x}}+\frac{\partial f/\partial p_{c,x}}{\sqrt{\pi^3}R_xR_yR_z}\nonumber\\
&\dot{p}_{c,x}=-\omega^2_x A_x \label{dipole}
\end{align}
where
\begin{eqnarray}
\frac{\partial}{\partial p_{c,x}}=\frac{1}{\cos^4\theta+\sin^4\theta}\left(\cos^2\theta \frac{\partial}{\partial p_{+,x}}+\sin^2\theta \frac{\partial}{\partial p_{-,x}}\right) \nonumber
\end{eqnarray}
The equation for breathing mode is the same as Eq. (\ref{breathing}). Moreover, for the stripe phase since $0<\theta<\pi/2$ and $\sin2\theta\neq 0$, there is one more dynamic equation for the relative phase $\varphi$ given by
\begin{align}
 \dot{\varphi}=\frac{1}{\sin2\theta}\frac{\partial \mathcal{E}}{\partial \theta}+(p_{+,x}-p_{-,x})\dot{A}_x \label{varphi}
\end{align}
Because the energy is independent of $\varphi$, there is no dynamics for $\theta$.

\begin{figure}[tbp]
\includegraphics[height=2.8in, width=3.2 in]
{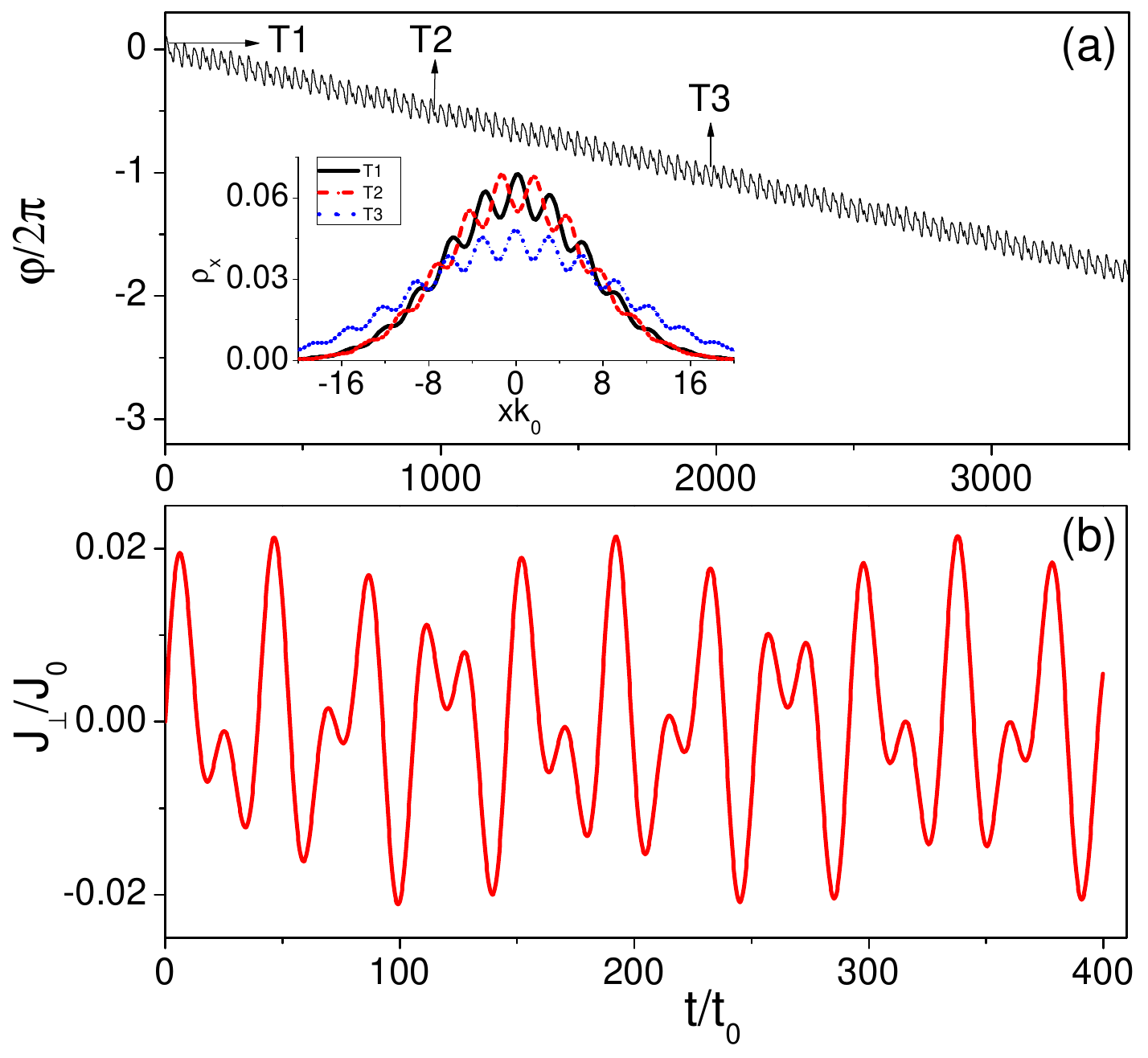}
\caption{Hall response in the stripe phase. (a) The relative phase $\varphi$ evolves as a function of $t$ after a sudden radial compression in the transverse direction. The inset illustrates the longitudinal spatial density profile $\rho_x=\int d^2r_{\perp}|\Phi|^2$ at different time $T_1$, $T_2$ and $T_3$, which demonstrates that the time evolution of $\varphi$ corresponds to sliding of density stripe. $x$ is in unit of $1/k_0$. (b) Time dynamics of radial current $J_{\perp}$ in the transverse direction after a sudden change of $\theta$. $t$ is in unit of $\hbar/2E_{\text{R}}$, and $E_{\text{R}}=\hbar^2 k^2_0/(2m)$. In both case, we choose parameters $\Omega=0.50 E_{\text{R}}$, $\delta=-0.24 E_{\text{R}}$, $\hbar\omega_x=0.33 E_{\text{R}}$ and $\hbar\omega_{\perp}=0.2 E_{\text{R}}$ which gives $\theta=0.85$, $k_0 R_x=12.6$ and $k_0 R_{\perp}=20.7$ for equilibrium phase. For (a), we suddenly compress $R_{\perp}$ to its half value, and for (b), we suddenly change $\theta$ to $1.05$.  \label{ST}}
\end{figure}

Here we discuss two types of Hall response shown in Fig. \ref{ST}. Note that for an equilibrium state, $\theta$ is given by the balance 
between the single particle energy and the interaction energy. First, in Fig. \ref{ST}a, 
if one excites a breathing mode along the transverse direction, 
this changes the interaction energy, and therefore $\partial \mathcal{E}/\partial \theta$ also becomes non-zero. The direct consequence is that $\varphi$ will change as a function of time according to Eq. (\ref{varphi}). This leads to the linear dependence of $\varphi$ on time $t$, as shown in Fig. \ref{ST}a. Moreover, the breathing mode also weakly couples to center-of-mass motion in the longitudinal direction $A_x$, thus, the second term in Eq. \ref{varphi} leads to small wiggles on top of the linear behavior. However, this effect is much weaker. Time evolution of $\varphi$ means that stripe slides along the longitudinal direction, as one can see from the inset of Fig. \ref{ST}a for density profile at different times. Since $\varphi$ corresponds to the phonon degree of density stripe, which is the most low-lying  excitation, and therefore it is the most dominative response in the stripe phase. Secondly, in Fig. \ref{ST}b, if we suddenly change $\theta$, it strongly alters the interaction energy and thus excites radial current in the transverse direction.

{\it Summary:} In summary, we have proposed two new types of Hall response in SO coupled condensates recently realized experimentally. These effects are remarkable because they are purely caused by interactions: (i) for the plane wave phase, because SO coupling locks spin to the velocity, the spin-dependent interaction energy changes as longitudinal velocity varies, and the change of interaction energy excites breathing motion along the transverse direction; and (ii) for the stripe phase, because the stripe order comes from the superposition of two wave functions with different momenta, and the weight of superposition $\sin\theta$(or $\cos\theta$) is determined by the interaction energy, thus the change of interaction energy by compressing the transverse confinement excites the dynamics of its conjugate variable $\varphi$, which describes the phonon mode of the stripe. As one can see, the first effect is directly connected to spin-velocity locking and the second one is connected to the nature of stripe phase, thus, although we use current experimental setup as an example to illustrate these two effects, they may also exist in a wide range of condensed matter and cold atom systems with SO coupling or stripe order.

{\it Acknowledgements.} This work is supported by Tsinghua University Initiative Scientific Research Program. HZ is supported by NSFC under Grant No. 11004118 and No. 11174176, and NKBRSFC under Grant No. 2011CB921500.

\end{document}